\documentstyle[prl,multicol,amsmath,aps,psfig]{revtex}
\begin{document}

\title{ 
Roughness at the depinning threshold for a long-range elastic string}
\author{Alberto Rosso and Werner Krauth
\footnote{rosso@lps.ens.fr; krauth@lps.ens.fr,
http://www.lps.ens.fr/$\tilde{\;}$krauth} 
}
\address{CNRS-Laboratoire de Physique Statistique \\
Ecole Normale Sup{\'{e}}rieure,
24, rue Lhomond, 75231 Paris Cedex 05, France}
\maketitle
\begin{abstract} 
In this paper, we compute the roughness exponent $\zeta$ of a long-range
elastic string,  at the depinning threshold,  in a random medium with
high precision, using a numerical method which exploits the analytic
structure of the problem (`no-passing' theorem), but avoids direct
simulation of the evolution equations.  This roughness exponent has
recently been studied by simulations, functional renormalization group
calculations, and by experiments (fracture of solids, liquid meniscus
in $^4$He). Our result $\zeta = 0.390 \pm 0.002$ is significantly larger
than what was stated in previous simulations, which were consistent with
a one-loop renormalization group calculation. The data are furthermore
incompatible with the experimental results for crack propagation in
solids and for a $^4$He contact line on a rough substrate. This implies
that the experiments cannot be described by pure harmonic long-range
elasticity in the quasi-static limit.

\end{abstract}

\begin{multicols}{2}
\narrowtext

The statics and dynamics of elastic manifolds in random media govern
the physics of a variety of systems, ranging from vortices in type-II
superconductors \cite{supra} and charge density waves \cite{CDW} to
interfaces in disordered magnets \cite{Lemerle}, contact lines of liquid
menisci on a rough substrate \cite{Prevost} and  to the propagation of
cracks in solids \cite{Gao}.

In most cases,  the restoring elastic forces  acting on a point on the
manifold are local {\em i.e.} depend only on the deformation in its
neighborhood.  The corresponding short-range string has been the object
of many theoretical and experimental studies. In the depinning limit, two
different scenarios are  possible: numerical simulations and analytical
calculations \cite{Jensen,Chauve} have established that a string with
an elastic restoring force   breaks at the depinning threshold, while
percolation experiments and numerical studies on directed polymers in
random media \cite{Buldyrev,RossoKrauth2} agree that in those systems
with stronger than harmonic  restoring forces the roughness exponent
is $\zeta=0.63$.

It has also been shown  \cite{Gao,Joanny} that for a contact line of
a liquid meniscus or for crack propagation in a solid, the elastic
force is long-range, rather than local.  Non-local elasticity  can be
expected to modify the dynamic and static properties of these systems
and to change the  critical exponents.  In this work,   we compute one
of these exponents, the  roughness exponent $\zeta$ of a long-range
elastic string  at the depinning threshold $f_c$.

The theoretical approaches  are up to  now based on the assumption
that  the motion of the line at the threshold is  quasi-static. This
means that velocity-dependent terms in the equations of motion of the
deformation field  $h(x,t)$ are taken to be irrelevant and can be derived
from an energy function, which incorporates potential energy due to the
driving force $f$ and the disorder potential $\eta(x,h)$, as well as an
elastic energy.  According to this hypothesis,
the equation of motion of the deformation field at zero temperature is:
\begin{equation}
\frac{ \partial  }{\partial t } h(x,t) 
=f + \eta(x,h) -k \int dx_{1}
\frac{h(x,t)-h(x_{1},t)}{(x-x_{1})^2}.
\label{motion}
\end{equation} 
The last term in this equation accounts for long-range restoring forces.
Let us note that measurements of local velocities for a liquid $^4$He
contact line \cite{Prevost}  have cast doubts on the validity of the
quasi-static hypothesis for the present experiments.

The critical behavior of eq.(\ref{motion}), at the driving force $f$
equal to $f_c$, was studied by  means of renormalization group (RG)
techniques.  The  one-loop calculations \cite{Ertas}  gave a roughness
exponent $\zeta^{(1)}$ equal to $1/3$  which at a time  was believed
to be exact \cite{Ertas,Narayan}.
Early simulations  based on extremal
long-range models \cite{Schmittbuhl1,Tanguy} ($\zeta=0.35 \pm 0.02$)
and on cellular automata \cite{Ramanathan} ($\zeta=0.34 \pm 0.02$) found
good agreement with this conjecture.
However, experiments, both  in crack propagation \cite{Schmittbuhl} and for
a liquid $^4$He contact line on a rough substrate \cite{Prevost}
have measured, near $f_c$, a systematically larger  exponent
($\zeta=0.56 \pm 0.03$).
Chauve {\sl et al.} \cite{Chauve} recently showed that 
higher-order  terms contribute to the RG result. At  two-loop
order, they found an exponent $\zeta^{(2)} =
0.47$. The large difference of the two-loop calculation with
the lowest-order result let it seem conceivable that the experimental
value could be explained by the model eq.(\ref{motion}).  In fact, the
authors of ref. \cite{Chauve} estimated the  exponent to be $\zeta=0.5
\pm 0.1$, which did include the experimental results.

In this article we present a new and very precise method to determine
the roughness exponent at $f_c$.  We show that the exponent $\zeta$
has  the value $0.390 \pm 0.002$. $\zeta$ is thus bigger than what was
suggested by earlier simulations ({\em cf}, however \cite{Zhou}) but is
at the same time incompatible with the experimental situation.

The direct numerical simulation of the dynamics, especially of 
long-range systems, in the depinning region is extremely tedious
because the velocity of the manifold vanishes at the threshold, which
is thus difficult to approach \cite{Marchetti,Leschhorn}.  Many authors
therefore preferred to treat the problem within the framework of
cellular automaton models.  These approaches can be very useful, but
are hardly identifiable with a continuum equation (concerning
the long-range case {\em cf} \cite{Ramanathan,Thomas}).

However, there is an additional analytical structure in this problem,
as first noticed by Middleton \cite{Middleton} for the continuum models.
For short-range lattice models \cite{RossoKrauth2,RossoKrauth1}, we
used this additional information to compute the critical string (the
blocked string at $f_c$) in an extremely efficient way, but without
actually simulating the time evolution of the system.  On the lattice,
the long-range model does not seem to be open to such an approach,
but the continuum model is, as we will show in this paper.

\begin{figure}
\centerline{ \psfig{figure=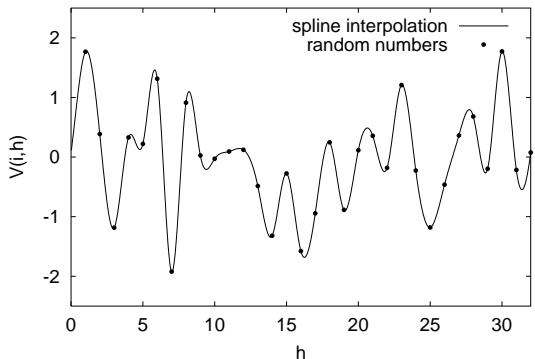,height=5.cm} }
\caption{ Example of a random potential on the $i$-th site with periodic
boundary conditions ($M=32$). The circles are Gaussian random numbers
with zero mean and unit variance, assigned to evenly spaced points
$h_i=1,2,\ldots,M$.  These  numbers are then interpolated  by a  periodic
cubic spline, in order to yield a continuous random potential $V(i,h_i)$.
}
\label{potential}
\end{figure}

To proceed, we discretize the variable $x$ ($x \rightarrow x_i,
i=1, \ldots, L$) in eq.(\ref{motion}) and write $h_i^t$ instead of
$h(x_i,t)$. The complete string at time $t$ is thus given by $h^t =
\{h_1^t,h_2^t,\ldots,h_L^t\}$, where the $h_i^t$ are real periodic
variables for which $h_i+M $ and $h_i$ are identified.  Periodic boundary
conditions are also applied in $L$: $h_{L+i}=h_i$.  The equation of motion
(\ref{motion}) is adapted to accomodate the periodic boundary conditions.
We have computed the long-range force $f^{el}$ by summing over periodic
images, as others have  done before us \cite{Schmittbuhl1,Tanguy},
obtaining:
\begin{equation}
f^{el}[h_i]= 
\sum_{i_{1}=1}^{L-1}
\left(2{\psi}^{'} \left( i_1/L \right) -
 \pi^2/sin^2(\pi i_1/L)  \right)
{\frac{h_{i}^t-h_{i_{1}}^t}{L^2}}
\end{equation} 
where  ${\psi}= d\Gamma(x)/dx$ and $ \Gamma(x)$  is the gamma function.
Different calculations, without the sum of images, 
gave an identical result for $\zeta$.  A random potential  $V(i,h_i)$ acts
on each site of the string.  Our choice of the random potential is shown
in fig.\ref{potential}. It allows to obtain a differentiable potential
with $\eta(i,h_i)=-{ \partial  V(i,h_i) }/{\partial h_i} $.

The `no-passing' rule \cite{Middleton} establishes the following:
if two strings $h$ and $\tilde{h}$ do not cross at a given time $t$
(say, $h_i^t < \tilde{h}_i^{t} \;\; \forall i$), they will not cross
at any later time.  Another important property  of Middleton's theorem
states that if, at an initial time $t_{init}$, the velocities $v_i^t $
are non-negative for all $i$, they will remain so  for all later times
$t > t_{init}$.  It follows from this property that, once we have found
a forward moving string $h^{t_{init}}$, we can be sure that snapshots of
the string at later times will never cross. In fact, the strings $h^{t}$
for $t>t_{init}$ will form a non-crossing family with non-negative
velocities, which satisfies the following conditions:
\begin{equation} 
\begin{array}{llll} 
i)   &  h_i^t \ge h_i^{t'}\;\; \forall i & \mbox{for}  & t > t' > t_{init} \\ 
ii)  & h_i^t \rightarrow h_i^{t'}      
  & \mbox{for}  & t > t'> t_{init}, \;\;
                                         t \rightarrow t' \\
iii) &  v(h_i^t) \ge 0 \;\;\forall i     & \mbox{for}  & t > t_{init} \\ 
iv)  &  \frac{\partial}{\partial t}  h_i^t = v(h_i^t)
                                         & \mbox{from} &
                                     \mbox{eq.}(\ref{motion})
\end{array} 
\end{equation} 
The velocity $v(h_i^t)$ is given, as usual, by $v(h_i^t)=-\partial
E[h^t]/\partial h_i^t $ where $E[h^t]$ is the energy of the configuration
$h^t$.  In this context, we have made the following observation: if our
only aim is to obtain the critical string, rather than to simulate the
true time-behavior, it is sufficient to generate continuous non-crossing
families satisfying $i)$,  $ii)$, and $iii)$, without imposing $iv)$. In
this case, $t$ would not be the physical time, but simply an ordering
index; only then are the three conditions $i)$, $ii)$, and $iii)$
independent.

\begin{figure} \centerline{ \psfig{figure=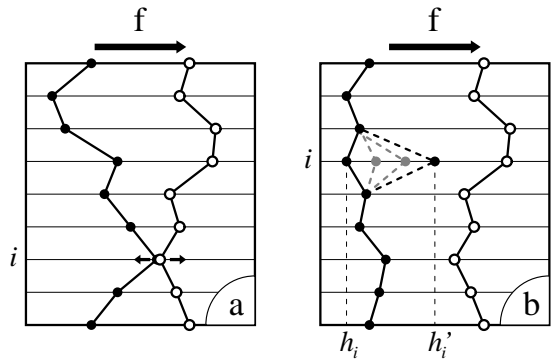,height=5.cm} }
\caption{ a: `No-passing' theorem:  $h^{t^*}$ (filled circles) and
$h^{block}$ (open circles). If $h^{t^*}$ approaches $h^{block}$  in
a point $i$, the disorder forces become equal, but the elastic term
(represented by arrows) prevents  $h^{t}$ from exceeding $h^{block}$
for $t>t^*$.  b: Algorithm. The point $i$ of the string (filled circles)
is moved from $h_i$ to $h'_i$ in one step. Between $h_i$ and $h_i'$,
the string's velocity in $i$ remains positve and and is zero at  $h'_i$.
A blocked configuration is drawn (open circles): our string may approach
this configuration, but  cannot pass it.}
\label{nopassing} 
\end{figure}
To assure that we can drop the condition $iv)$, we have to guarantee that
no member of such a family will ever cross a blocked string $h^{block}$
(by definition $v(h_i^{block})\equiv 0$), if $h^{t_{init}}$ did not.
Let us suppose the contrary:  if $h^{t^*} $ were the first  member
which   touches the blocked line in one point  $i$ (as shown in the
fig.(\ref{nopassing}a)), the random force for $h_i^{block}$ and $h^{t^*}_i
$ would become equal, but  the elastic  term would give $v(h^{t^*}_i)<
v(h^{block}) $. This would be  a  violation of condition $iii)$.
Conversely, it is easy to see that such a family can always be continued
up to a blocked line, because it suffices to find a single point $i$
with positive velocity to continue the construction.

In previous works \cite{Marchetti}, the critical line was computed by
direct simulation of eq.(\ref{motion}).  Notice that the discretization of
time can pose difficult problems: during the  interval time  $\Delta t$
the motion does not respect Middleton's theorem and therefore is not
guaranteed to halt in front of a blocked string.

The construction of continuous non-crossing families of strings presents
a much more powerful,  completely rigorous approach.  As indicated in
fig.(\ref{nopassing}), we consider in practice families in which at
a given instant only a single coordinate $i$ moves.  Coordinate $i$
is then advanced (from $h_i$ to $h'_i$) until its velocity vanishes
($v(h'_i)=0$).  This is of course not done by simulation, but in a single
step, by computing the zeros of $v(h_i)$.  To simplify our numerical
work, we have  represented the continuous random potential as a spline
(a piecewise third-order polynomial in $h$), as further explained in
fig.(\ref{potential}). This allows us to solve for the closest zero of
the velocity function from a quadratic equation.  The only parameter in
this procedure is the minimal velocity below which the {\em total} string
is assumed blocked, and which serves to terminate our iteration.  We have
varied this non-essential parameter by four orders of magnitude, and shown
that the critical forces and exponents are extremely well stabilized.

An initial forward-moving string $h^{t_{init}}$ is very easy to obtain.
We have remarked that our iterative algorithm, which is completely
rigorous, converges extremely quickly to a blocked string if such an
object exists.  A bisection method then allows us to obtain the critical
driving force $f_c$, and the critical string.  We stress again that,
by construction, a  blocked string is never passed.

We have run this algorithm on a large number of samples 
with $L=M$. It can be shown that
the correct thermodynamic limit is to consider change of scale as $(L,M)
\rightarrow (\alpha L, \alpha M)$ \cite{RossoKrauth1}.  Sample sizes
varied from $L=64 $ to $L= 1024$, where we are able to obtain a critical
string (at $f=f_c$) in a few  minutes on a PC.  We compute the mean-square
elongation $W^2(L)$ of a critical string $h^c$ as
\begin{equation}
W^2(L): = \overline{
 < (h^c -   <h^c>)^2>
}.
\label{elongdef}
\end{equation}
In eq.(\ref{elongdef}), $<h^c> = \frac{1}{L}\sum_i h^c_i$, while the
overbar stands for an average over the disorder.  Our data obviously
extrapolate very well  and a mean-square analysis yields 
$\zeta =0.390 \pm 0.002$.
\begin{figure}
\centerline{ \psfig{figure=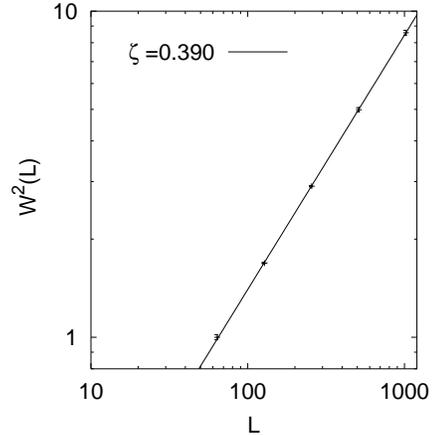,height=6.cm} }
\caption{Mean square elongation $W^2(L)$ as a function of system size
$L$ for the  long-range elastic force given by eq.(\ref{motion}).
The interpolating line corresponds to a roughness exponent $\zeta = 0.390
\pm 0.002$. 
 }
\label{main}
\end{figure}
In conclusion, we have obtained the roughness exponent for long-range
elastic strings by a numerical procedure which respects the analytic
structure in the problem (`no-passing' theorem) and allows to obtain
very high precision. Our approach can certainly be extended to other
problems, we would in particular   be interested in simulation methods
which allowed to study, {\em e.g.}, the KPZ equation \cite{KPZ} directly in
the continuum.  This is most important, as it has been shown that the
effect of discretizations can be very difficult to control \cite{Newman}.

The large difference of the result with the experimental value shows
that the theoretical model of these processes will certainly have to be
modified  in an essential way. One possibility is that velocity-dependent
terms have to be taken into account. This appears reasonable, as large
local velocities have been observed in the experiments.

Acknowledgements: 
We would like to thank P.~Le~Doussal, S.~Moulinet and E.~Rolley for very
helpful discussions.

\end{multicols}

\end{document}